\documentclass[aps,nofootinbib]{revtex4}

\newcommand{\lan}{\langle}
 \newcommand{\ran}{\rangle}
 \newcommand{\be}{\begin{equation}}
 \newcommand{\bea}{\begin{eqnarray}}
 \newcommand{\eea}{\end{eqnarray}}
 \newcommand{\ee}{\end{equation}}

\def\fun#1#2{\lower3.6pt\vbox{\baselineskip0pt\lineskip.9pt
\ialign{$\mathsurround=0pt#1\hfil##\hfil$\crcr#2\crcr\sim\crcr}}}

\newcommand{\eff}{\mathop{\rm eff}\nolimits}
\newcommand{\dia}{\mathop{\rm dia}\nolimits}

\begin{document}
\title{Chiral thermodynamics in a magnetic field }

\author{Nikita O. Agasian}
\email{agasian@heron.itep.ru}
\affiliation{Institute of Theoretical and Experimental Physics,\\
Moscow 117218, Russia}
\date{December 27, 2001}

\begin{abstract}
We study thermodynamic properties of the QCD vacuum in a magnetic field
below chiral phase transition. The hadronic phase free energy in a
constant homogeneous  magnetic field is calculated in the framework of
the chiral perturbation theory at non-zero pionic mass. It is demonstrated
that the  order parameter of the chiral phase transition remains constant
provided temperature and magnetic field strength are related through
obtained equation (the phenomenon of ''quark condensate freezing'').
\end{abstract}

\maketitle

%\newpage

\section{Introduction}
  The investigation of the vacuum state behavior under the
 influence of the various external factors is known to be one of the
 central problems in quantum field theory. In the realm of strong
 interactions (QCD) the main factors are the temperature and
 the baryon density. At
temperatures below the chiral phase transition, $T<T_c$, the dynamics
of the system is characterized by confinement and spontaneous
breaking of chiral symmetry (SBCS). At low temperatures, $T<T_c$,
the partition function of the system is dominated by the contribution
 of the lightest  particles in the physical spectrum. In QCD this
role is played by the $\pi $ -- meson which is the Goldstone
excitation mode in chiral condensate. Therefore the low
temperature physics (the hadron phase) enables an adequate
description in terms of the effective chiral theory \cite{1,2,3}.
A very important problem is the behavior of the order parameter
(the quark condensate $\langle \bar q q\rangle)$ with the increase
of the temperature.  In the ideal gas approximation the
contribution of the thermal pions into the quark condensate is
proportional to $ T^2$
 \cite{4,5}. In chiral perturbation theory (ChPT) the two -- and
three--loops contributions ($\sim T^4$  and  $\sim T^6$
 correspondingly) into $\langle \bar q q \rangle$ have been found in
\cite{5}  and \cite{6,6a}.
At non-zero quark mass the analytic temperature dependence of the
quark condensate was found in \cite{ag3} in perfect agreement with the numerical
calculations obtained at the three-loop level of the chiral
perturbation theory with $m_q\neq0$ \cite{6a}.

The situation  with the gluon condensate
 $\langle G^2\rangle\equiv \langle
(G^a_{\mu\nu})^2\rangle$is very different. The gluon condensate
is not an order parameter in phase transition and it does not lead
to any spontaneous symmetry breaking (SSB). At the quantum level
the trace anomaly leads to the breaking of the scale invariance
and  this in turn results in nonzero value of
 $\langle G^2 \rangle$.  However, this is not a SSB phenomenon and
hence does not lead to the appearance of the Goldstone particle. The
mass of the lowest excitation (dilaton)  is directly connected to the
gluon condensate, $ m_D\propto
 (\langle G^2\rangle)^{1/4}$.
Thus the thermal excitations of glueballs are exponentially
suppressed by the Boltzmann factor $\sim \exp
\{-m_{gl}/T\}$ and their contribution to the shift of the gluon
condensate is small ($\Delta\langle G^2\rangle/\langle G^2\rangle\sim
$0.1 \% at $T=$200 MeV) \cite{7.a}.  Next we note that in the
one-loop approximation of ChPT pions are described as a gas of
massless noninteracting particles. Such a system is obviously
scale-invariant and therefore does not contribute into the trace of
the  energy-momentum tensor and correspondingly into $\langle
G^2\rangle$.
It was shown in \cite{7} that only in the three-loop approximation
of ChPT does the gluon condensate become dependent on temperature in the chiral limit.
At non-zero quark mass the analytic temperature dependence of the
gluon condensate was found in \cite{ag3}.
The effect of thermal excitations of massive hadrons on the properties
of the quark and the gluon condensate was studied in \cite{agebil} within
the generalized nonlinear $\sigma$-model invariant under conformal transformations.

Another interesting problem is the study of the vacuum phase
structure under the influence of the external magnetic field $H$.
Quarks play an active role in shaping the QCD vacuum structure.
Being dual carriers of both 'color' and 'electric' charges they
also respond to externally applied electromagnetic fields. The
vacuum of strong interactions influences some QED  processes has
been discussed in  Ref. \cite{Raf}. The behavior of the quark
condensate in the presence of a magnetic field was studied in
Nambu-Iona-Lasinio model earlier \cite{0.8}. For QCD, the
analogous investigation in the one-loop approximation of ChPT was done in
\cite{8}. It was found that the quark condensate grows with the
increase of the magnetic field $H$ in both cases. It implies that
a naive analogy with superconductivity, where the order parameter
vanishes at same critical field, is not valid here. The behavior
of the gluon condensate $\langle G^2\rangle$ in the Abelian
magnetic field is also a nontrivial effect. Gluons do not carry
electric charge; nevertheless, virtual quarks produced by them
interact with electromagnetic field and lead to  the changes in
the gluon condensate.  This phenomenon was studied in \cite{10},
\cite{11} based on the low-energy theorems in QCD. The vacuum
energy density,
 the values of  $\langle G^2\rangle$  and $\langle \bar q q\rangle$
as functions of $H$ have been found in the two--loop approximation
of ChPT in \cite{11}.

The low-energy theorems, playing an important role in the
understanding of the vacuum state properties in quantum field
theory, were discovered almost at the same time as quantum field
methods appeared in particle physics (see, for example Low
theorems \cite{12}).In QCD, these theorems were obtained in the
beginning of eighties \cite{13}.  These theorems, being derived
from the very general symmetrical considerations and not depending
on the details
 of confinement mechanism, sometimes give information which is not
easy to obtain in another way. Also, they can be used as
"physically sensible" restrictions in the constructing of
effective theories. An important step  was made in \cite{14},
where low-energy theorems for  gluodynamics were generalized
to finite temperature case.
A relation between the trace anomaly and thermodynamic pressure in pure-glue
QCD was obtained in \cite{Land} by making use of the dimensional regularization in
the framework of the renormalization group (RG) method. Also within the RG method,
but by employing slightly different techniques an analogous relation was
derived in the theory with quarks in  \cite{ag1}.

The QCD phase structure in a magnetic field at finite temperature
was investigated in \cite{ag2}. Within the ChPT framework the
dependence of the quark condensate on $T$ and $H$ was studied
and it was shown that the shift of the condensate is not the
simple sum of the temperature ($\sim T^2/F^2_\pi$) and
magnetic ($\sim H/(4 \pi F_\pi)^2$) contributions.
There appears an additional term which physically stems
from the orbital diamagnetism of the charged pion gas.
In the chiral limit the thermodynamic quantities, in
particular the quark condensate, are characterized by additional
dimensionless parameter $\lambda=\sqrt{H}/T$.
However as it was shown in \cite{agsh} the Gell-Mann-Oakes--Renner relation
is not changed in a magnetic field at $T \neq 0$ (no additional
terms appear) and thus the soft chiral symmetry breaking
scenario remains the same.

It should be noted, that ablian type chromomagnetic fields in QCD
are essential for the nonperturbative dynamics of vacuum. The role of
chromomagnetic fields in the thermal deconfining phase was investigated
in~\cite{sim}. The influence of chromomagnetic vacuum condensate on
the phenomenon of color superconductivity was studied in~\cite{CSC}.
The rearrangement of quark-antiquark vacuum in the strong gluonic fields
was studied in~\cite{agvosk}.

In the present paper the vacuum free energy in magnetic field at
finite temperature is calculated in the framework of ChPT. The
general relations are established which allow to obtain the
dependence of the quark and gluon condensates on $T$ and $H$. A new
phenomenon is displayed, namely the "freezing" of the chiral phase
transition order parameter by the  magnetic field when the
temperature increases. The physical meaning of this fact  is
discussed.

\section{Renormalization group properties of the gluon condensate at
$T\neq 0$ and $H\neq 0$}
 For non-zero quark mass ($m_q\neq 0)$ and in a magnetic field
 the scale invariance is
 broken already at the classical level.
Therefore the pion thermal  excitations would change, even in the
ideal gas approximation, the value of the gluon condensate with
increasing temperature and field \footnote{ At zero quark mass
and in the absence of magnetic field
 the gas of massless noninteracting pions is obviously
 scale-invariant and therefore does not contribute to
the trace of the  energy-momentum tensor and correspondingly to
the gluon condensate $\langle (G^a_{\mu\nu})^2\rangle$.}.
 To determine this dependence use will be made of the general
 renormalization and scale properties of the QCD partition
 function.

 The QCD Euclidean partition function  with two quark flavors
 in  external Abelian field $A_\mu$
 has the following form ($T=1/\beta$)
 \be
 Z=exp \left \{ -\frac{1}{4e^2}
\int^\beta_0 dx_4\int_V d^3x F^2_{\mu\nu} \right \}
 \int[DB][D\bar q][Dq]
\exp \left \{ -\int^\beta_0 dx_4\int_V d^3x {\cal L} \right \}.
\label{eq_1}
 \ee
  Here the QCD Lagrangian in the background field is
\be
 {\cal L}=\frac{1}{4g^2_0}
 (G^a_{\mu\nu})^2+ \sum_{q=u,d} \bar q[\gamma_\mu
 (\partial_\mu-iQ_q A_\mu-i\frac{\lambda^a}{2} B^a_\mu)+m_{0q}]q,
 \label{eq_2}
  \ee
where $Q_q$ -- is the matrix of the quark charges for the quarks
$q=(u,d)$, and for the simplicity the ghost terms have been omitted.

 The free energy density is given by the relation $ \beta VF$
$(T,H,m_{0u},m_{0d})=-\ln Z$.  Eq. (\ref{eq_2}) yields
the following expression for the gluon condensate
($\langle G^2\rangle\equiv \langle (G^a_{\mu\nu})^2\rangle$)
  \be
    \langle G^2\rangle (T,H,m_{0u},m_{0d})=4\frac{\partial F}{\partial(1/g^2_0)}~.
 \label{eq_4}
 \ee
 The system described by the partition function (\ref{eq_2}) is
 characterized by the set of dimensionful parameters $M, T, H, m_{0q}
 (M)$ and dimensionless charge $g^2_0(M)$, where $M$ is the
 ultraviolet cutoff.
 On the other hand one can consider the renormalized free energy
 $F_R$ and by using the dimensional and renormalization-group properties
 of $F_R$ recast (\ref{eq_4}) into the form containing derivatives
 with respect to the physical parameter $T$, $H$ and renormalized masses $m_q$.

The phenomenon of dimensional transmutation results in the
appearance of a nonperturbative dimensionful parameter
 \be
  \Lambda= M \exp
 \left \{ \int^\infty_{\alpha_s(M)}\frac{d\alpha_s}{\beta(\alpha_s)}
 \right \}~,
  \label{eq_5}
  \ee
  where
  $\alpha_s=g^2_0/4\pi$, and $\beta(\alpha_s)=d\alpha_s(M)/d
  ~ln M$  is the Gell-Mann-Low  function.
Furthermore, as it is well known,    the quark mass has anomalous
dimension and depends on the scale $M$. The  renormalization
-group equation for $m_0(M)$, the running mass, is  $d\ln m_0/d\ln
M=-\gamma_m$ and we use the $\overline{MS} $ scheme for which
$\beta$ and $\gamma_m$ are independent of the quark mass
\cite{muta}. Upon integration the renormalization-group invariant
mass is given by
\be
m_q=m_{oq}(M)\exp\{\int^{\alpha_s(M)}\frac{\gamma_{m_q}(\alpha_s)}{\beta(\alpha_s)}
d\alpha_s\}~,
 \label{eq_6}
 \ee
  where the indefinite integral is
evaluated at $\alpha_s(M)$. Next we note that since free energy is
renormalization-group invariant quantity its anomalous dimension
is zero. Thus $F_R$ has only a normal (canonical) dimension equal
to 4. Making use of the renorm-invariance of $\Lambda$, one can
write in the most general form
\be
F_R=\Lambda^4 f(\frac{T}{\Lambda}, \frac{H}{\Lambda^2}, \frac{m_u}{\Lambda},
\frac{m_d}{\Lambda})~,
 \label{eq_7}
 \ee
 where $f$ is some function.
 From (\ref{eq_5}),(\ref{eq_6}) and (\ref{eq_7}) one gets
\be
\frac{\partial F_R}{\partial(1/g^2_0)}=
 \frac{\partial
F_R}{\partial\Lambda} \frac{\partial\Lambda}{\partial(1/g^2_0)} +
\sum_q \frac{\partial F_R}{\partial m_q} \frac{\partial
m_q}{\partial(1/g^2_0)}~,
 \label{eq_8}
 \ee

\be \frac{\partial m_q}{\partial(1/g^2_0)}=-4\pi\alpha^2_s
m_q\frac{\gamma_{m_q}(\alpha_s)}{\beta(\alpha_s)}~.
 \label{eq_9}
 \ee
 With the account of (\ref{eq_4}) the gluon condensate is given by
\be
\lan G^2\ran (T, H, m_u, m_d)
=\frac{16\pi\alpha_s^2}{\beta(\alpha_s)}(4-T\frac{\partial}{\partial
T}-2 H\frac{\partial}{\partial H}-\sum_q(1+\gamma_{m_q})m_q\frac{\partial}{\partial {m_q}}) F_R.
\label{eq_10_0}
 \ee
 It is convenient to choose such a large scale that one can take the
lowest order expressions,  $\beta(\alpha_s)\to -
b\alpha^2_s/2\pi$, where $b=(11 N_c-2N_f)/3$ and $1+\gamma_m\to
1$. Thus, we have the following equations for condensates
\be
\lan G^2\ran (T, H, m_u, m_d)=-\frac{32\pi^2}{b} (4-T\frac{\partial}{\partial
T}-2 H\frac{\partial}{\partial H}-\sum_q m_q\frac{\partial}{\partial m_q}) F_R\equiv
 -\hat DF_R~,
 \label{eq_11_0}
 \ee
 \be
 \lan\bar q q\ran (T, H, m_u, m_d)=\frac{\partial F_R}{\partial {m_q}}~.
 \label{eq_12_0}
 \ee

\section{Free energy of the QCD vacuum at low temperature
in a magnetic field}

The above equations enable to obtain the values of the
condensates as functions of
 $T$  and  $H$ provided the free energy density is known. To get the
latter the ChPT will be used. At low temperatures
 $T<T_c (T_c$ is the chiral phase transition temperature) and for weak
fields
 $H< \mu^2_{hadr} \sim (4\pi F_\pi)^2$ the characteristic momenta
in the vacuum loops are small and theory is adequately described
by the
        low-energy effective
 chiral Lagrangian
 $L_{\eff}$ \cite{2,3}.
This Lagrangian can be represented as a series expansion  over
the momenta (derivatives) and quark masses
 \be
L_{\eff}=L^{(2)}+L^{(4)}+L^{(6)}+...
\label{eq_10}
 \ee
 The leading term in
(\ref{eq_10}) is similar to the Lagrangian of the non-linear sigma model
in the external field
 $$ L^{(2)}=\frac{F^2_\pi}{4}Tr(\nabla_\mu
U^+\nabla_\mu U)+\Sigma Re~Tr(\hat M U^+),
$$
\be
\nabla_\mu
U=\partial_\mu U-i[U,V_\mu].
 \label{eq_11}
  \ee
  Here $U$ is a unitary
$SU(2)$  matrix, $F_\pi=93 $MeV  is the pion decay constant, and
 $\Sigma$  has the meaning of the quark condensate
 $\Sigma =|\langle \bar u u\rangle | = |\langle
\bar d d\rangle |$.  The external Abelian magnetic field $H$ is aligned
 along the  $z$ -axis and corresponds to $V_\mu(x)=(\tau^3/2) A_\mu(x)$
 with the vector-potential $A_\mu$  chosen as $A_\mu(x)=\delta_{\mu 2}Hx_1$.
The mass difference between the
 $u$ and  $d$ quarks appears in the effective chiral Lagrangian
 only quadratically. Further, to obtain an expression for the
quark condensate in the chiral limit we use only the first
 derivative with respect to the mass of one of the quarks.
Therefore, we can neglect the mass difference between the $u$ and $d$
quarks and assume the mass matrix to be diagonal $\hat M=m\hat I$.

At  $T<T_c, H< \mu^2_{hadr}$ the QCD partition function coincides
with the partition function of the effective chiral theory
 \be
  Z_{\eff}[T,H]=e^{-\beta
VF_{\eff}[T,H]}=Z_0[H]\int [DU] \exp \{ -\int^\beta_0 dx_4\int_Vd^3x
L_{\eff} [U,A]\}
 \label{eq_12}
  \ee
  At the one-loop level it is sufficient to restrict the expansion
of $L_{\eff}$  by the quadratic terms with respect to the pion
field. Using the exponential parameterization of the matrix $U(x)=
\exp \{ i\tau^a\pi^a(x) /F_\pi\} $ one finds
\be L^{(2)}=\frac12 (\partial_\mu
\pi^0)+\frac12 M^2_\pi(\pi^0)^2+ (\partial_\mu\pi^++iA_\mu\pi^+)
(\partial_\mu\pi^--iA_\mu\pi^-)   + M^2_\pi \pi^+\pi^-,
\label{eq_13}
\ee
where  the charged $\pi^\pm$  and neutral  $\pi^0$  meson fields are introduced
 \be
\pi^\pm =(\pi^1\pm i\pi^2)/\sqrt{2},~~ \pi^0=\pi^3
 \label{eq_14}
  \ee
Thus  (\ref{eq_12})  can be recasted into the form \footnote{ The
partition function $Z_{\eff}^R$ describes charged $\pi^{\pm}$ and
neutral $\pi^0$ ideal Bose gas in magnetic field. Relativistic
charged Bose gas in magnetic field at finite temperature and
density with application to Bose-Einstein condensation and Meissner
effect was studied in Refs.  \cite{Tom}, \cite{Elm},
\cite{Roj}.}
\be Z_{\eff}^R[T,H]=Z^{-1}_{p.t.} Z_0[H]\int [D\pi^0]
[D\pi^+][D\pi^-] \exp \{-\int^\beta_0 dx_4\int_Vd^3x L^{(2)}[\pi,
A]\}
\label{eq_15}
\ee
    where partition  function is normalized for the case of
    perturbation theory  at
     $T=0, H=0$
\be Z_{p.t.} = [\det
    (-\partial^2_\mu+M^2_\pi)]^{-3/2}.
\label{eq_16.a}
\ee
     Integration of
    (\ref{eq_15}) over $\pi$-fields leads to
     \be
     Z_{\eff}^R=Z^{-1}_{p.t.}Z_0[H] [{{\rm
 det}_T(-\partial^2_\mu+M^2_\pi)}]^{-1/2} [{{\rm
 det}_T(-|D_\mu|^2+M^2_\pi)}]^{-1},
 \label{eq_17.a}
 \ee
 where
 $D_\mu=\partial_\mu-iA_\mu$ is a covariant derivative
 and a symbol  $"T"$  means that the determinant is calculated
 at finite temperature
  $T$ according to standard Matsubara rules. Taking
  (\ref{eq_16.a})  into account and regrouping multipliers in
  (\ref{eq_17.a}) one gets the following expression for $Z_{\eff}^R$
  $$ Z_{\eff}^R [T,H]=Z_0[H] \left[
 \frac{\det_T(-\partial^2_\mu+M^2_\pi)}
   {\det(-\partial^2_\mu+M^2_\pi)}\right ] ^{-1/2}
  \left[
   \frac{\det
 (-|D_\mu|^2+M^2_\pi)}
   {\det(-\partial^2_\mu+M^2_\pi)}\right ] ^{-1}
  $$
 \be
 \times \left[
   \frac{\det_T
 (-|D_\mu|^2+M^2_\pi)}
   {\det(-|D_\mu|^2+M^2_\pi}\right ] ^{-1}
 \label{eq_18.a}
 \ee
 Then the effective free energy can be written in the form
  \be
  F_{\eff}^R(T,
 H)=-\frac{1}{\beta V}\ln Z_{\eff}^R
 =\frac{H^2}{2e^2}+F_{\pi^0}(T)+F_{\pi^{\pm}}(H) +F_{\rm dia}(T,H).
 \label{eq_16}
 \ee
 Here  $F_{\pi^0}$  is the free energy of massive scalar boson
 \be
F_{\pi^0}(T)=T\int\frac{d^3p}{(2\pi)^3}\ln (1-\exp (-\sqrt{{\bf
p}^{2}+M^2_\pi}/T)),
 \label{eq_17}
 \ee
 $F_{\pi^\pm}$ is a Schwinger result for the vacuum energy
density of charged scalar particles in the magnetic field
 \be
F_{\pi^\pm}(H)=-\frac{1}{16\pi^2}\int^\infty_0\frac{ds}{s^3}
e^{-M^2_\pi s}  [\frac{Hs}{\sinh (Hs)}-1],
\label{eq_18}
\ee
and $F_{\rm dia}$ is the diamagnetic free energy of relativistic
charged Bose gas
$$
F_{\rm dia}(T,H)=\frac{HT}{\pi^2}\sum^\infty_{n=0}\int^\infty_0 dk\ln
(1-\exp(-\omega_n/T)),
$$
\be \omega_n=\sqrt{k^2+M^2_\pi+H(2n+1)},
\label{eq_19}
\ee where
$\omega_n$ are Landau levels of the  $\pi^\pm$ mesons in constant
field $H$. \footnote{Technically, a transition
 for the free energy
  $F=\frac12 Tr \ln
 (p^2_4+\omega^2_0(\bf p))$
 from the vacuum case ($H=0, T=0)$ to the case of  $ H\neq 0, T\neq
 0$
 is straightforward.
  Omitting the details of
  the calculations, we note that, eventually, this
transition reduces to the substitutions
  $p_4\to \omega_k=2\pi kT$
    ($k=0,\pm
 1,...$), $\omega_0=\sqrt{{\bf p}^2+M^2_\pi}
 \to \omega_n=\sqrt{p^2_z+M^2_\pi+H(2n+1)}
 $
  and
 $Tr\to \frac{HT}{2\pi} \sum^\infty_{n=0}
\sum^{+\infty}_{k=-\infty}\int^{+\infty}_{-\infty}
\frac{dp_z}{2\pi},$
 where the degeneracy multiplicity of  $H/2\pi$ has been
 taken into
 account for the Landau levels. Performing summation over Matsubara
 frequencies, we obtain (\ref{eq_19}).}

 By expanding integrand in~(\ref{eq_17}), (\ref{eq_19}) in the series, one obtains
 the following expressions

  \be
  F_{\pi^0}=-\frac{M^2_\pi T^2}{2\pi^2}\sum^\infty_{n=1} \frac{1}{n^2}
  K_2(n\frac{M_\pi}{T})
  \label{eq_fpi0}
  \ee
and
  \be
  F_{\rm dia}=-\frac{HT}{\pi^2}
   \sum^\infty_{n=0}
   \sqrt{M^2_\pi+H(2n+1)}
   \sum^\infty_{k=1}\frac{1}{k}
   K_1\left ( \frac{k}{T}
   \sqrt{M^2_\pi+H(2n+1)}\right)
    \label{eq_fdia}
   \ee
where $K_n$ is the Mackdonald function.

\section{Quark condensate at $T \neq 0$ and $H \neq 0$}

The free energy  $F_{\eff}^R$  determines the thermodynamical
properties and the phase structure of the QCD vacuum state below the
temperature of the chiral phase transition, i.e. in the phase  of
confinement.

 In order to get the dependence of the quark condensate upon
  $T$ and  $H$
 use is made of the Gell-Mann-Oakes--Renner relation
  \be
  F^2_\pi
 M^2_\pi=-\frac12(m_u+m_d)\langle\bar u u +\bar d d\rangle =2m \Sigma
 \label{eq_fpi}
 \ee
The quark condensate shift is then determined by
\be
\frac{\Delta\Sigma(T,H,M_\pi)}{\Sigma}=
-\frac{1}{F^2_\pi}\frac{\partial F^R_{\eff}}{\partial
M^2_\pi}
\label{eq_shift}
\ee
Expression for quark condensate as function of $T$ and $H$ at $M_{\pi}\neq 0$
is then given by
\be
\frac{\langle \bar{q}q\rangle (T,H,M_\pi)}{\langle \bar{q}q\rangle} =
1+\frac{\Delta{\Sigma}_{\pi^0}(T,M_\pi)}{\Sigma}+
\frac{\Delta{\Sigma}_{\pi^{\pm}}(H,M_\pi)}{\Sigma} +
\frac{\Delta{\Sigma}_{\rm {dia}}(T,H,M_\pi)}{\Sigma}
\label{eq_qq}
\ee
where the contribution of thermal $\pi^0$-meson to the shift of $\langle\bar{q}q\rangle$ is
  \be
  \frac{\Delta
  \Sigma_{\pi^0}}{\Sigma} =-\frac{TM_\pi}{4\pi^2F^2_\pi}\sum^\infty_{n=1}
  \frac{1}{n} K_1(n\frac{M_\pi}{T}),
  \label{eq_sigma_pi0}
  \ee
vacuum contribution of $\pi^{\pm}$-mesons (Shwinger term) is
   \be
   \frac{\Delta \Sigma_{\pi^\pm}}{\Sigma} =-\frac{H}{(4\pi F_\pi)^2}
   \int^\infty_{0}\frac{dx}{x^2}
   \left
   [\frac{x}{\sinh(x)}-1\right ]
   \exp\left(-\frac{M^2_\pi}{H}\, x\right)
   \label{eq_sigma_pipm}
   \ee
and ''diamagnetic contribution'' of charged Bose gas of $\pi^{\pm}$-mesons
to the quark condensate is given by
   \be
   \frac{\Delta \Sigma_{\rm dia}}{\Sigma} =-\frac{H}{2\pi^2 F_\pi^2}
   \sum^\infty_{n=0}
   \sum^\infty_{k=1}
   K_0\left ( \frac{k}{T}
   \sqrt{M^2_\pi+H(2n+1)}\right)
    \label{eq_sigma_dia}
   \ee

In the limiting case of weak magnetic field $H \ll M_{\pi}^2$ and low
temperature $T \ll M_{\pi}$ one gets
   \be
   \frac{\Delta \Sigma_{\pi^0}}{\Sigma}=-\sqrt{8\pi}
   \frac{M^{1/2}_\pi T^{3/2}}{(4\pi F_\pi)^2} e^{-M_\pi/T} +
   O(e^{-2M_{\pi}/T})
   \label{eq_sigma_pi0_2}
   \ee
and
   \be
   \frac{\Delta \Sigma_{\pi^{\pm}}}{\Sigma}=
   \frac{H^2}{96\pi^2F^2_\pi M^2_\pi} +
   O(\frac{H^4}{M_{\pi}^4})
   \label{eq_sigma_pipm_2}
   \ee

Let us now consider diamagnetic term $\Delta\Sigma_{\dia}$ for
various limiting cases.
In the case of weak magnetic field,
the interval between Landau levels $\sim \sqrt{H}$ is much less
then average thermal energy of $\pi^{\pm}$-mesons,
$\sqrt{H} \ll T$, and we can use
Euler-MacLaren summation formula. We then find
for $\Delta\Sigma_{\dia}$
\be
\frac{\Delta\Sigma_{\dia}}{\Sigma}=2\frac{\Delta\Sigma_{\pi^0}}{\Sigma}+
\frac{H^2}{48 \pi^2 F_{\pi}^2 M_{\pi} T} \sum_{n=1}^{\infty} n K_1 \left(
n \frac{M_{\pi}}{T}\right),\quad \sqrt{H} \ll T
\label{eq_sigma_dia_2}
\ee

For the case of low temperature $T \ll M_{\pi}$ one obtains
\be
\frac{\Delta\Sigma_{\dia}}{\Sigma}=2\frac{\Delta\Sigma_{\pi^0}}{\Sigma}+
\frac{H^2}{48 \sqrt{2} \pi^{3/2} F_{\pi}^2 M^{3/2}_{\pi} T^{1/2}} e^{-M_{\pi}/T},
\quad \sqrt{H} \ll T \ll M_{\pi}
\label{eq_sigma_dia_3}
\ee

In the opposite limit of low temperature and strong magnetic field (at $M_{\pi}\neq 0$)
asymptotic behavior of $\Delta \Sigma_{\dia}$ is determined by $n=0$, $k=1$ in
the sum~(\ref{eq_sigma_dia})

\be
\frac{\Delta\Sigma_{\dia}}{\Sigma}=-\frac{H}{2\pi^2 F^2_\pi} K_0\left (
\frac{\sqrt{M^2_\pi+H}}{T}\right )
\to -\frac{1}{(2\pi)^{3/2} F^2_\pi}\frac{H T^{1/2}}{(M^2_\pi+H)^{1/4}}
e^{-\sqrt{M^2_\pi+H}/T}, \quad T \ll \sqrt{H} \ll M_{\pi}
\label{eq_sigma_dia_asymp}
\ee

   An interesting
  phenomenon reveals itself in the vacuum QCD phase structure  under
  consideration. One can find from (\ref{eq_qq}) such a function $H(T)$
  that the chiral condensate $\langle \bar q q\rangle (T,H,M_{\pi})$ remains
  unchanged when the temperature and magnetic field change in
  accordance with $H_*=H(T)$. Then $H_*$ is found by solving the
  following equation  $\langle \bar q q\rangle
  (T,H_*,M_{\pi})- \langle \bar q q\rangle=0$ (see (\ref{eq_qq})). At low temperature,
  $T \ll M_{\pi}$,
  and weak field, $\sqrt{H} \ll M_{\pi}$, we have

\be
H_*(T,M_{\pi})=\sqrt{24}\left( \frac{\pi}{2}\right)^{1/4} M_{\pi}^{7/4} T^{1/4}
e^{-{M_{\pi}}/{2T}}
+O(e^{-{M_{\pi}}/{T}})
\label{eq_Hstar}
\ee

Let us now consider quark condensate as function of $T$ and $H$ in chiral limit $M_{\pi}=0$
~\cite{ag2}.

 Substituting (\ref{eq_16}) into (\ref{eq_shift}), calculating the derivative
 over $M^2_\pi$ and then taking the limit $M^2_\pi\to 0$ one gets
 $$ \langle \bar q q\rangle (T,H)= \langle \bar q
 q\rangle (1-\frac13\cdot \frac{T^2}{8F^2_\pi}+\frac{H}{(4\pi
 F_\pi)^2} \ln 2-\frac{H}{2\pi^2F^2_\pi} \varphi(\frac{\sqrt{H}}{T}))
 $$
 \be
\varphi(\lambda)=\sum^\infty_{n=0}\int^\infty_0\frac{dx}{\omega_n(x)
(\exp(\lambda\omega_n(x))-1)}, ~~ \omega_n(x)=\sqrt{x^2+2n+1}
\label{eq_22} \ee Now we consider various limiting cases. In the
strong  field, $\sqrt{H} \gg T$ $(\lambda \gg 1)$, the lowest
Landau level ($n=0$) gives the main contribution to the sum
(\ref{eq_22})
\be
\varphi(\lambda\gg 1) = \sqrt{\frac{\pi}{2\lambda}}e^{-\lambda}
  +O(e^{-\sqrt{3}\lambda}).
  \label{eq_23}
  \ee
   In the opposite limit of weak field, $\sqrt{H}\ll T(\lambda\ll 1)$, the
  sum in (\ref{eq_22}) is calculated with required accuracy using
  the Euler-MacLaren formula. Furthermore, one gets the following result with the use of
  the  asymptotic expansion of integral (\ref{eq_22})\cite{Kap} at $\lambda\ll 1$
  \be
  \varphi (\lambda\ll 1) =\frac{\pi^2}{6} \frac{1}{\lambda^2}
  +\frac{7\pi}{24} \frac{1}{\lambda} +\frac{1}{4} \ln \lambda
  +C+\frac{ \zeta(3)}{48\pi^2} \lambda^2 +O(\lambda^4),
  \label{eq_24}
  \ee
   here $C=\frac14(\gamma-\ln 4\pi-\frac16),~~ \gamma =0.577...$
   is Euler's constant and $\zeta(3)=1.202$ is Riemann zeta function.
  Thus, one obtains the following limiting expressions for the quark
  condensate in the chiral limit in a magnetic
  field at $T\neq 0$
  \be
  \frac{\langle \bar qq\rangle (T,H)}{\langle \bar q
  q\rangle}=1-\frac13\cdot \frac{T^2}{8F^2_\pi}
  +\frac{H}{(4\pi F_\pi)^2} \ln2-\frac{H^{3/4}T^{1/2}}{(2\pi)^{3/2}
  F^2_\pi} e^{-\sqrt{H}/T},~~\sqrt{H}\gg T
  \label{eq_25}
  \ee
  and
  \be
  \frac{\langle \bar qq\rangle (T,H)}{\langle \bar q
  q\rangle}=1-\frac{T^2}{8F^2_\pi}
  +\frac{H}{(4\pi F_\pi)^2}
  A-\frac{7 \sqrt{H}T}{48 \pi F_\pi^2}
  - \frac{H}{(4\pi F_\pi)^2}
  \ln\frac {H}{T^2},
  ~~\sqrt{H}\ll T
  \label{eq_26}
  \ee
  where
  $A=\ln 2-8C\simeq 4.93$.

   In the framework of ChPT the quark condensate (\ref{eq_22}) at
   $H\neq 0$, $T\neq 0$ is determined by three dimensionless parameters
   $H/(4\pi F_{\pi})^2$, $T^2/F_{\pi}^2$ and $\lambda=\sqrt{H}/T$.
   The quantity $\lambda$ is a natural dimensionless parameter in
   this approach. The motion of a particle (massless pion) in the
   field is characterized by the curvature radius of it's
   trajectory, and in the magnetic field this is the Larmor radius
   $R_L=1/\sqrt{H}$. On the other hand, there is another length $l_T=1/T$ -
   "temperature length" at $T\neq 0$.
   Therefore, charged $\pi^{\pm}$-mesons in
   magnetic field effectively acquire "mass", $m_{\eff} = \sqrt{H}$,
   determined by the lowest Landau level, when Larmor radius of a
   particle in the field is much less than $l_T (\lambda \gg 1)$.
   Correspondingly, their
   contribution to the shift of the chiral condensate is
   suppressed by the Boltzman factor $\propto\exp\{-m_{\eff} / T\}$.
   In the weak field limit $\pi^{\pm}$-mesons give
   standard temperature one-loop approximation ChPT contribution
   to $\langle \bar q q\rangle$. Besides, additional temperature
   and magnetic corrections appear.
   Neutral $\pi^0$-meson contributes to
   $\langle \bar q q\rangle(T,H)$ as usual massless scalar
   particle.

  In the chiral limit the equation for $H_*(T)$  takes the form
   \be
   1-\frac{3}{2\pi^2} \lambda^2\ln 2
  +\frac{ 12}{\pi^2}\lambda^2\varphi(\lambda)=0
  \label{eq_27}
  \ee
  The numerical
  solution of  (\ref{eq_27})  yields $\lambda_*=3.48...$ Thus, quark
  condensate stays unchanged when $T$ and $H$ are increased according
  to $H=12.11\cdot T^2$.  Hence it is possible to say that the order
  parameter $\langle \bar q q\rangle$  of the chiral phase transition
  is "frozen" by the magnetic field.

  Note that $H(T_c)/ (4\pi F_\pi)^2\simeq 0.2 \ll 1$ at $T=T_c\simeq 150 $MeV
and therefore the above relations remain valid up to the deconfined phase
transition point \footnote{There was an arithmetical mistake in paper~\cite{ag2},
and the value of $\lambda_*$ was underestimated. However, this mistake does not affect
any physical consequences.}.
In the vicinity of $T_c$ the effective low energy chiral
Lagrangian fails to provide an adequate description of the QCD vacuum
thermodynamical properties, and strictly speaking becomes physically invalid.
  The following is worth noting. In deriving
(\ref{eq_22}), at the first step the physical quantity
 as functions of $M_\pi$ where obtained, and
only then the chiral limit
 $M_\pi\to
0$ was taken. Acting in the inversed sequence we would have
obtained all temperature corrections to condensate identically
equal to zero. This points to the fundamental difference of the
two cases: the exactly massless particle and the particle with
infinitesimal small mass.

\section{Conclusion}
 It has been shown in the present letter that the quark
condensate is "frozen" by the magnetic field when both temperature
$T$ and magnetic field $H$ are increased according to the
obtained relation $H=H_*(T,M_{\pi})$. This points to the fact that the direct analogy
  between the quark condensate in QCD and the theory of
  superconductivity is untenable. In the BCS theory the Cooper pairs
  condensate is extinguished by the temperature and magnetic field.
  The "freezing" phenomenon can be understood in terms of the
  general Le Chatelier--Braun principle \footnote{The external action
  disturbing the system from the equilibrium state induces processes
  in this system which tend to reduce the result of this action}.The
  external field contributes into the system an additional energy
  density
  $H^2/2$. The system tends to compensate this energy change and to
  decrease the free energy by
  increasing the absolute value of the quark condensate:
  $\Delta\varepsilon_v=-m|\Sigma(H)-\Sigma(0)|<0$. On the other hand,
  if the temperature of the system is increased (by bringing some
  heat into it) the processes with heat absorbtion
  by damping the condensate are switched on. The interplay of these
  processes is at the origin of the above "freezing" of $\Sigma
  (T,H)$.

  In $N_f=2$ QCD, the temperature phase transition restoring chiral
  $SU(2)_L \times SU(2)_R$ symmetry is a second-order phase transition.
  As the temperature is increased, the order parameter $\Sigma(T)$ decreases
  monotonically, vanishing at the critical point $T_c$. The existence of the
  jump $\Delta \Sigma(T_c)\neq 0$ would indicate the occurrence of a firs-order
  phase transition. The order parameter at any non-zero magnetic field
  $H$ is greater than that at $H=0$; hence, $\Sigma(T,H)-\Sigma(T,H=0)>0$. Because
  of this, it may turn out that the chiral phase transition in a magnetic
  field becomes a first-order phase transition, and critical temperature
  of phase transition is larger at $H \neq 0$, $T_c(H)>T_c(H=0)$. In order
  to clarify the character of the phase transition, it is necessary to investigate
  the behavior of the system and of the order parameter in the fluctuation
  region near $T_c$; however, the effects of $\pi\pi$ interaction are substantial
  in this region. It follows that, within the approach used here, it is
  impossible to establish the change in the order of transition, although it seems
  highly probable \footnote{The presence of an external field (not necessarily a
  magnetic field) transforms a second-order phase transition into a first-order
  phase transition~\cite{landau}}. This phenomenon may prove to be of use in
  investigating various cosmological scenarios after the Big Bang. For this reason,
  it would be interesting to continue studying the chiral phase transition in a
  magnetic field.

\section*{ACKNOWLEDGMENTS}

The author is grateful to B.L. Ioffe, V.A. Novikov, V.A. Rubakov, Yu.A. Simonov and
A.V. Smilga  for comments and discussions.
The financial support of RFFI grant 00-02-17836 and
INTAS grant N 110 is gratefully acknowledged.

\end{document}